# Version 5
# Structured psychosocial stress and the US obesity epidemic

Rodrick Wallace, Ph.D.
The New York State Psychiatric Institute
Deborah N. Wallace, Ph.D.
Joseph L. Mailman School of Public Health
Columbia University*

December 8, 2003

## Abstract

We examine the accelerating 'obesity epidemic' in the US from the perspective of generalized language-of-thought arguments relating a cognitive hypothalamic-pituitary-adrenal axis to an embedding context of structured psychosocial stress. From a Rate Distortion perspective, the obesity epidemic is an image of ratcheting social pathology – indexed by massive, policy-driven, deurbanization and deindustrialization – impressed upon the bodies of American adults and children. The resulting pattern of developmental disorder, while stratified by expected divisions of class and ethnicity, is nonetheless relentlessly engulfing even affluent majority populations.

## Introduction

Obesity is epidemic in the United States, has been so for more than two decades, and continues to increase. Current rates of overweight and obesity in the US are 61% and 14% in adults and children respectively. Obesity in adults has nearly doubled since 1980, from 15% in 1980 to 27% by 1999 (e.g. Wellman and Friedberg, 2002). Childhood overweight is rapidly rising in the US, particularly among African Americans and Hispanics. By 1998 prevalence increased to 21.5% in African Americans, 21.8% among Hispanics, and 12.3% among non-Hispanic whites aged 4 to 12 years (Strauss and Pollack, 2001).

The obesity epidemic is associated with serious health conditions including type 2 diabetes, heart disease, high blood pressure and stroke, certain types of cancer, hypoxia, sleep apnea, hernia, and arthritis. It is a major cause of economic loss, death, and suffering which shows no indications of abatement.

A recent series of articles in *Science* (Vol. 299, 7 February, 2003) examines the problem from an individualized perspective which largely fails to explore the epidemiology and population ecology of the problem. The piece by Hill et al. (2003), for example, explains the obesity epidemic as the simple disjunction between calorie intake and output consequent on eased workload, 'larger food portions', and disinclinations to exercise. In some contrast, our perspective places the 'explanation' that 'obesity occurs when people eat too much and get too little exercise' in the same category as the remark by US President Calvin Coolidge on the eve of the Great Depression that 'unemployment occurs when large numbers of people are out of work'. Both statements ignore profound structural determinants of great population suffering which must be addressed by collective actions of equally great reform.

Indeed, experts on health disparities have long recognized that obesity is unevenly distributed geographically, ethnically, and by socioeconomic class. Urban people of color (Allan, 1998), poor Southern states (Mokdad et al., 1999), and poor neighborhoods within cities (Ginsberg-Fellner et al, 1981) have higher prevalences. The Southern states form the epicenter of the geographically spreading epidemic (Mokdad et al., 1999), a picture of contagion between populations.

The famous Whitehall Studies of British civil servants (Brunner et al., 1997) found that coronary heart disease and central abdominal fat deposition incidences were strongly associated with the occupational hierarchy. Locus of work control was a major factor in both central abdominal fat deposition and coronary heart disease. Power relations in the workplace imposed a particular structure of stress.

Furthermore, stress which causes sleep deficits shifts metabolism toward fat accumulation and central abdominal deposition (Spiegel et al, 1999). The Hypothalamic-pituitary-adrenal (HPA) axis is central to the mechanisms (Chrousos, 2000; Bjorntorp, 2001). So the stress involves adrenal reactions to serious threats.

Our hypothesis, in contrast to that of the *Science* special issue, is that large numbers of Americans feel seriously threatened. The obesity epidemic embodies the consequences of public policies: economic insecurity from deindustrialization, social upheaval from destruction of cities by programs of 'benign neglect' and 'planned shrinkage', (Wallace and Wallace, 1998), the nation's wealth increasingly concentrated in fewer and fewer hands, and a voting ritual which, as the last US presidential election shows clearly, doesn't seem to matter. The US obesity epidemic embodies, in our view, a worsen-ing crisis of democratic locus-of-control which will not be addressed by platitudes about 'eating less and exercising more'.

*Address correspondence to R. Wallace, PISCS Inc., 549 W. 123 St., Suite 16F, New York, NY, 10027. Telephone (212) 865-4766, rdwall@ix.netcom.com. Affiliations are for identification only.

## Stress and the HPA axis

Abdominal obesity and visceral fat accumulation are particularly associated with disease, and have become the focus of much research on 'stress' and its relation to the 'fight-or-flight' responses of the HPA axis. We paraphrase Bjorntorp (2001), who extensively summarizes the role of the HPA axis in physiological responses to stress.

When the input of noxious signals is prolonged, the HPA axis reactivity changes from normal and relatively transient attempts to maintain homeostasis or allostasis with temporary peaks of cortisol secretion, first, to a state of sensitization, which reacts with exaggerated cortisol secretion after a given input. This occurs during the most active phase of the HPA axis, which is the early morning in humans. When repeated too often and with sufficient strength of the input, the first sign of malfunction is a delayed down-winding, so that cortisol secretion stays elevated for a prolonged period of time. Subsequently, the normal diurnal pattern is disrupted, and morning values tend to be lower. This subsequently develops into a low, steady, rigid diurnal cortisol secretion with little reactivity, a 'burned out' HPA.

In parallel, the controlling, central glucocorticoid receptors become less efficient, and down-regulated. Further challenges are followed by atrophy of the entire system, often found after long-term, severe hypercortisolaemia as in Cushing's syndrome, melancholic depression, post traumatic stress disorder (PTSD), and the aftermath of war.

Bjorntorp (2001) describes how elevation of cortisol is followed by visceral fat accumulation. Much research shows consequent lowered sex steroid and growth hormone secretions have the same consequence, because of the insufficient counteraction against cortisol effects, and the combination of these abnormalities powerfully directs a larger than normal fraction of total body fat to visceral deposits.

In sum, increased activity of the HPA axis triggers both inhibition of both the pituitary gonadal and growth hormone axes. Stress may, then, synergistically cause accumulation of visceral fat, via elevated cortisol secretion and decrease of sex steroid and growth hormones.

Bjorntorp concludes in particular that the deposit of central body fat, which is closely correlated with general measures of obesity, can serve as a reasonable approximation to the long-term endocrine abnormalities associated with stress and often-repeated or chronic activation of the HPA axis.

That is, stress literally writes an image of itself onto the body as visceral fat accumulation, first having written an image of itself onto the HPA axis. The phenomenon can be interpreted as the transmission of a structured signal between communicating systems, in a large sense, i.e. psychosocial to HPA.

Here we will adapt recent developments regarding the punctuated information dynamics of evolutionary process to the question of how the communication of the embedding psychosocial structure and the HPA axis might be constrained by certain of the asymptotic limit theorems of probability. We know that, regardless of the distribution of a particular stochastic process, the Central Limit Theorem ensures that long sums of independent realizations of that process will follow a Normal distribution. Similar constraints exist on the behavior of communicating structures, and are described by the limit theorems of information theory. Importation of phase transition methods from statistical physics, done much in the spirit of the Large Deviations Program of applied probability, permits concise and unified description of evolutionary and cognitive 'learning plateaus' which, in the evolutionary case, are interpreted as evolutionary punctuation (e.g. Wallace, 2002a, b).

'Stress', we aver, is not often random in human societies, but is rather a highly structured 'language', having both a grammar and a syntax, so that certain stressors are 'meaningful' in a particular context, and others are not, having little or no long-term physiological effect. We will argue that the HPA axis is, in fact, a cognitive system itself associated with a 'dual information source' which may also be expressed as a kind of language. It is the punctuated interaction of these two 'languages' which we will find critical to an understanding of how psychosocial stress affects the HPA axis, and, ultimately, writes a distorted image of itself on the human body as visceral fat deposition.

This analysis presents a slightly different picture of the obesity epidemic, but one having profound implications for intervention policy.

## HPA axis cognition

Atlan and Cohen (1998) argue that the essence of cognition is comparison of a perceived external signal with an internal, learned picture of the world, and then, upon that comparison, the choice of a response from a much larger repertoire of possible responses. Clearly, from this perspective, the HPA axis, the 'flight-or-fight' reflex, is cognitive. Upon recognition of a new perturbation in the surrounding environment, memory and brain cognition evaluate and choose from several possible responses: no action necessary, flight, fight, helplessness (i.e. flight or fight needed, but not possible). Upon appropriate conditioning, the HPA axis is able to accelerate the decision process, much as the immune system has a more efficient response to second pathogenic challenge once the initial infection has become encoded in immune memory. Certainly 'hyperreactivity' in the context of PTSD is a well known example. This is almost certainly an evolutionary adaptation of considerable significance.

Following the approach of Wallace (2000, 2002a), we make a very general model of that process.

Pattern recognition-and-response, as we characterize it, proceeds by convoluting (i.e. comparing) an incoming external 'sensory' signal with an internal 'ongoing activity' – the 'learned picture of the world' – and, at some point, triggering an appropriate action based on a decision that the pattern of sensory activity requires a response. We need not model how the pattern recognition system is 'trained', and hence we adopt a weak model, regardless of learning paradigm, which can itself be more formally described by the Rate Distortion Theorem. We will, fulfilling Atlan and Cohen's (1998) criterion of meaning-from-response, define a language's contextual meaning entirely in terms of system output.

The model is as follows.

A pattern of sensory input is convoluted (compared) with internal 'ongoing' activity to create a path of convoluted signal $x = (a_0, a_1, ..., a_n, ...)$. This path is fed into a highly nonlinear 'decision oscillator' which generates an output $h(x)$ that is an element of one of two (presumably) disjoint sets $B_0$ and $B_1$. We take

$$B_0 \equiv b_0, ..., b_k,$$

$$B_1 \equiv b_{k+1}, ..., b_m.$$

Thus we permit a graded response, supposing that if

$$h(x) \in B_0$$

the pattern is not recognized, and if

$$h(x) \in B_1$$

the pattern is recognized and some action $b_j, k+1 \leq j \leq m$ takes place.

We are interested in paths $x$ which trigger pattern recognition-and-response exactly once. That is, given a fixed initial state $a_0$, such that $h(a_0) \in B_0$, we examine all possible subsequent paths $x$ beginning with $a_0$ and leading exactly once to the event $h(x) \in B_1$. Thus $h(a_0, ..., a_j) \in B_0$ for all $j < m$, but $h(a_0, ..., a_m) \in B_1$.

For each positive integer $n$ let $N(n)$ be the number of paths of length $n$ which begin with some particular $a_0$ having $h(a_0) \in B_0$ and lead to the condition $h(x) \in B_1$. We shall call such paths 'meaningful' and assume $N(n)$ to be considerably less than the number of all possible paths of length $n$ – pattern recognition-and-response is comparatively rare. We further assume that the finite limit

$$H \equiv \lim_{n \to \infty} \frac{\log[N(n)]}{n}$$

both exists and is independent of the path $x$. We will – not surprisingly – call such a pattern recognition-and-response cognitive process *ergodic*.

We may thus define an ergodic information source $\mathbf{X}$ associated with stochastic variates $X_j$ having joint and conditional probabilities $P(a_0, ..., a_n)$ and $P(a_n|a_0, ..., a_{n-1})$ such that appropriate joint and conditional Shannon uncertainties may be defined which satisfy the relations

$$H[\mathbf{X}] = \lim_{n \to \infty} \frac{\log[N(n)]}{n} =$$

$$\lim_{n \to \infty} H(X_n|X_0, ..., X_{n-1}) =$$

$$\lim_{n \to \infty} \frac{H(X_0, ..., X_n)}{n}.$$

(1)

We say this information source is *dual* to the ergodic cognitive process.

The $H$-functions are cross-sectional sums of terms typically having the form

$$-\sum_k P_k \log[P_k]$$

where the $P_k$ are joint or conditional probabilities, so that equation (1) represents a kind of law-of-large-numbers for 'language' systems. See Ash (1990) or Cover and Thomas (1991) for details.

Different 'languages' will, of course, be defined by different divisions of the total universe of possible responses into different pairs of sets $B_0$ and $B_1$, or by requiring more than one response in $B_1$ along a path. Like the use of different distortion measures in the Rate Distortion Theorem (e.g. Cover and Thomas, 1991), however, it seems obvious that the underlying dynamics will all be qualitatively similar. Dividing the full set of possible responses into the sets $B_0$ and $B_1$ may itself, however, require 'higher order' cognitive decisions by other modules. This is an important matter to which we will return below, and serves as the foundation for speculations regarding obesity as a developmental disorder.

Meaningful paths – creating an inherent grammar and syntax – are defined entirely in terms of system response, as Atlan and Cohen (1998) propose.

We can apply this formalism to the stochastic neuron in a neural network: A series of inputs $y_i^j, i = 1, ...m$ from $m$ nearby neurons at time $j$ to the neuron of interest is convoluted with 'weights' $w_i^j, i = 1, ..., m$, using an inner product

$$a_j = \mathbf{y}^j \cdot \mathbf{w}^j \equiv \sum_{i=1}^m y_i^j w_i^j$$

(2)

in the context of a 'transfer function' $f(\mathbf{y}^j \cdot \mathbf{w}^j)$ such that the probability of the neuron firing and having a discrete output $z^j = 1$ is $P(z^j = 1) = f(\mathbf{y}^j \cdot \mathbf{w}^j)$.

Thus the probability that the neuron does not fire at time j is just $1 - P$. In the usual terminology the $m$ values $y_i^j$ constitute the 'sensory activity' and the $m$ weights $w_i^j$ the 'ongoing activity' at time $j$, with $a_j = \mathbf{y}^j \cdot \mathbf{w}^j$ and the path $x \equiv a_0, a_1, ..., a_n, ...$. A little more work leads to a standard neural network model in which the network is trained by appropriately varying $\mathbf{w}$ through least squares or other error minimization feedback. This can be shown to replicate rate distortion arguments, as we can use the error definition to define a distortion function which measures the difference between the training pattern $y$ and the network output $\hat{y}$ as a function, for example, of the inverse number of training cycles, $K$. As we will discuss in another context, 'learning plateau' behavior emerges naturally as a phase transition in the parameter $K$ in the mutual information $I(Y, \hat{Y})$.

Thus we will eventually parametize the information source uncertainty of the dual information source to a cognitive pattern recognition-and-response with respect to one or more variates, writing, e.g. $H[\mathbf{K}]$, where $\mathbf{K} \equiv (K_1, ..., K_s)$ represents a vector in a parameter space. Let the vector $\mathbf{K}$ follow some path in time, i.e. trace out a generalized line or surface $\mathbf{K}(t)$. We will, following the argument of Wallace (2002b), assume that the probabilities defining $H$, for the most part, closely track changes in $\mathbf{K}(t)$, so that along a particular 'piece' of a path in parameter space the information source remains as close to memoryless and ergodic as is needed for the mathematics to work. Between pieces we impose phase transition characterized by a renormalization symmetry, in the sense of Wilson (1971). See Binney, et al. (1986) for a more complete discussion.

We will call such an information source 'adiabatically piecewise memoryless ergodic' (APME). The ergodic nature of the information sources is a generalization of the 'law of large numbers' and implies that the long-time averages we will need to calculate can, in fact, be closely approximated by averages across the probability spaces of those sources. This is no small matter.

The reader may have noticed parallels in this argument with the large body of speculation on the language-of-thought hypothesis, or, more generally, what Adams (2003) has called 'the informational turn in philosophy'. Our innovation is to return attention from 'the information content of individual symbols' to the asymptotic properties of long streams of symbols, a kind of thermodynamic limit allowing importation of phase transition and quasi-thermodynamic methods from statistical physics via a homology between information source uncertainty and free energy density.

Note that our treatment does not preclude the existence of cognitive processes or submodules which may not have appropriately simple dual information sources. We cannot, however, address them in any obvious way, although some generalization may be possible to a subset of non-ergodic sources (e.g. Wallace, 2003).

**Interacting information sources**

We suppose that the behavior of the HPA axis can be represented by a sequence of 'states' in time, the 'path' $x \equiv x_0, x_1, ...$. Similarly, we assume an external signal of 'structured psychosocial stress' can be similarly represented by a path $y \equiv y_0, y_1, ...$. These paths are, however, both very highly structured and, within themselves, are serially correlated and can, in fact, be represented by 'information sources' $\mathbf{X}$ and $\mathbf{Y}$. We assume the HPA axis and the external stressors interact, so that these sequences of states are not independent, but are jointly serially correlated. We can, then, define a path of sequential pairs as $z \equiv (x_0, y_0), (x_1, y_1), ...$.

The essential content of the Joint Asymptotic Equipartition Theorem is that the set of joint paths $z$ can be partitioned into a relatively small set of high probability which is termed *jointly typical*, and a much larger set of vanishingly small probability. Further, according to the JAEPT, the *splitting criterion* between high and low probability sets of pairs is the mutual information

$$I(X,Y) = H(X) - H(X|Y) = H(X) + H(Y) - H(X,Y)$$

(3)

where $H(X), H(Y), H(X|Y)$ and $H(X,Y)$ are, respectively, the Shannon uncertainties of $X$ and $Y$, their conditional uncertainty, and their joint uncertainty. See Cover and Thomas (1991) for mathematical details. Similar approaches to neural process have been recently adopted by Dimitrov and Miller (2001).

The high probability pairs of paths are, in this formulation, all equiprobable, and if $N(n)$ is the number of jointly typical pairs of length $n$, then

$$I(X,Y) = \lim_{n \to \infty} \frac{\log[N(n)]}{n}.$$

(4)

Generalizing the earlier language-on-a-network models of Wallace and Wallace (1998, 1999), we suppose there is a 'coupling parameter' $P$ representing the degree of linkage between the immune system's reset cognition and the system of external signals and stressors, and set $K = 1/P$, following the development of those earlier studies. Then we have

$$I[K] = \lim_{n \to \infty} \frac{\log[N(K,n)]}{n}.$$

The essential 'homology' between information theory and statistical mechanics lies in the similarity of this expression with the infinite volume limit of the free energy density. If $Z(K)$ is the statistical mechanics partition function derived from the system's Hamiltonian, then the free energy density is determined by the relation

$$F[K] = \lim_{V \to \infty} \frac{\log[Z(K)]}{V}.$$

(5)

$F$ is the free energy density, $V$ the system volume and $K = 1/T$, where $T$ is the system temperature.

We and others argue at some length (e.g. Wallace and Wallace, 1998, 1999; Rojdestvensky and Cottam, 2000; Feynman, 1996) that this is indeed a systematic mathematical homology which, we contend, permits importation of renormalization symmetry into information theory. Imposition of invariance under renormalization on the mutual information splitting criterion $I(X,Y)$ implies the existence of phase transitions analogous to learning plateaus or punctuated evolutionary equilibria in the relations between the cognitive mechanism of the HPA axis and the system of external perturbations. An extensive mathematical treatment of these ideas is presented elsewhere (e.g. Wallace, 2000, 2002a,b; Wallace et al., 2003).

Elaborate developments are possible. From a the more limited perspective of the Rate Distortion Theorem we can view the onset of a punctuated interaction between the cognitive mechanism of the HPA and external stressors as a distorted image of those stressors within the HPA axis:

Suppose that two (piecewise, adiabatically memoryless) ergodic information sources $\mathbf{Y}$ and $\mathbf{B}$ begin to interact, to 'talk' to each other, i.e. to influence each other in some way so that it is possible, for example, to look at the output of $\mathbf{B}$ – strings $b$ – and infer something about the behavior of $\mathbf{Y}$ from it – strings $y$. We suppose it possible to define a retranslation from the B-language into the Y-language through a deterministic code book, and call $\hat{\mathbf{Y}}$ the translated information source, as mirrored by $\mathbf{B}$.

Define some distortion measure comparing paths $y$ to paths $\hat{y}$, $d(y,\hat{y})$ (Cover and Thomas, 1991). We invoke the Rate Distortion Theorem's mutual information $I(Y,\hat{Y})$, which is the splitting criterion between high and low probability pairs of paths. Impose, now, a parametization by an inverse coupling strength $K$, and a renormalization symmetry representing the global structure of the system coupling.

Extending the analyses, triplets of sequences can be divided by a splitting criterion into two sets, having high and low probabilities respectively. For large $n$ the number of triplet sequences in the high probability set will be determined by the relation (Cover and Thomas, 1992, p. 387)

$$N(n) \propto \exp[nI(Y_1;Y_2|Y_3)],$$

(6)

where splitting criterion is given by

$$I(Y_1;Y_2|Y_3) \equiv$$

$$H(Y_3) + H(Y_1|Y_3) + H(Y_2|Y_3) - H(Y_1,Y_2,Y_3)$$

We can then examine mixed cognitive/adaptive phase transitions analogous to learning plateaus (Wallace, 2002b) in the splitting criterion $I(Y_1,Y_2|Y_3)$. Note that our results are almost exactly parallel to the Eldredge/Gould model of evolutionary punctuated equilibrium (Eldredge, 1985; Gould, 2002).

Note that the expression above can be generalized to a number of interacting information sources, $Y_j$, embedded in a larger context, $Z$, as

$$I(Y_1,...,Y_s|Z) = H(Z) + \sum_{j=1}^{s} H(Y_j|Z) - H(Y_1,...,Y_s,Z)$$

### The simplest HPA axis model

Stress, as we envision it, is not a random sequence of perturbations, and is not independent of its perception. Rather, it involves a highly correlated, grammatical, syntactical process by which an embedding psychosocial environment communicates with an individual, particularly with that individual's HPA axis, in the context of social hierarchy. We view the stress experienced by an individual as APME information source, interacting with a similar dual information source defined by HPA axis cognition.

Again, the ergodic nature of the 'language' of stress is essentially a generalization of the law of large numbers, so that long-time averages can be well approximated by cross-sectional expectations. Languages do not have simple autocorrelation patterns, in distinct contrast with the usual assumption of random perturbations by 'white noise' in the standard formulation of stochastic differential equations.

Let us suppose we cannot measure stress, but can determine the concentrations of HPA axis hormones and other biochemicals according to some 'natural' time frame, which we will characterize as the inherent period of the system. Suppose, in the absence of extraordinary 'meaningful' psychosocial stress, we measure a series of $n$ concentrations at time $t$ which we represent as an $n$-dimensional vector $X_t$. Suppose we conduct a number of experiments, and create a regression model so that we can, in the absence of perturbation, write, to first order, the concentration of biomarkers at time $t+1$ in terms of that at time $t$ using a matrix equation of the form

$$X_{t+1} \approx <\mathbf{R}> X_t + b_0,$$

(7)

where $<\mathbf{R}>$ is the matrix of regression coefficients and $b_0$ a vector of constant terms.

We then suppose that, in the presence of a perturbation by structured stress

$$X_{t+1} = (<\mathbf{R}> + \delta\mathbf{R}_{t+1})X_t + b_0$$

$$\equiv <\mathbf{R}> X_t + \epsilon_{t+1},$$

(8)

where we have absorbed both $b_0$ and $\delta\mathbf{R}_{t+1}X_t$ into a vector $\epsilon_{t+1}$ of 'error' terms which are not necessarily small in this formulation. In addition it is important to realize that this is not a population process whose continuous analog is exponential growth. Rather what we examine is more akin to the passage of a signal – structured psychosocial stress – through a distorting physiological filter.

If the matrix of regression coefficients $<\mathbf{R}>$ is sufficiently regular, we can (Jordan block) diagonalize it using the matrix of its column eigenvectors $\mathbf{Q}$, writing

$$\mathbf{Q}X_{t+1} = (\mathbf{Q}<\mathbf{R}>\mathbf{Q}^{-1})\mathbf{Q}X_t + \mathbf{Q}\epsilon_{t+1},$$

(9)

or equivalently as

$$Y_{t+1} = <\mathbf{J}> Y_t + W_{t+1},$$

(10)

where $Y_t \equiv \mathbf{Q} X_t$, $W_{t+1} \equiv \mathbf{Q} \epsilon_{t+1}$, and $<\mathbf{J}> \equiv \mathbf{Q} <\mathbf{R}> \mathbf{Q}^{-1}$ is a (block) diagonal matrix in terms of the eigenvalues of $<\mathbf{R}>$.

Thus the (rate distorted) writing of structured stress on the HPA axis through $\delta \mathbf{R}_{t+1}$ is reexpressed in terms of the vector $W_{t+1}$.

The sequence of $W_{t+1}$ is the rate-distorted image of the information source defined by the system of external structured psychosocial stress. This formulation permits estimation of the long-term steady-state effects of that image on the HPA axis. The essential trick is to recognize that because everything is (APM) ergodic, we can either time or ensemble average both sides of equation (10), so that the one-period offset is absorbed in the averaging, giving an 'equilibrium' relation

$$<Y> = <\mathbf{J}><Y> + <W>$$

or

$$<Y> = (\mathbf{I} - <\mathbf{J}>)^{-1} <W>,$$

(11)

where $\mathbf{I}$ is the $n \times n$ identity matrix.

Now we reverse the argument: Suppose that $Y_k$ is chosen to be some fixed eigenvector of $<\mathbf{R}>$. Using the diagonalization of $<\mathbf{J}>$ in terms of its eigenvalues, we obtain the average excitation of the HPA axis in terms of some eigentransformed pattern of exciting perturbations as

$$<Y_k> = \frac{1}{1 - <\lambda_k>} <W_k>$$

(12)

where $<\lambda_k>$ is the eigenvalue of $<Y_k>$, and $<W_k>$ is some appropriately transformed set of ongoing perturbations by structured psychosocial stress.

The essence of this result is that *there will be a characteristic form of perturbation by structured psychosocial stress – the $W_k$ – which will resonantly excite a particular eigenmode of the HPA axis.* Conversely, by 'tuning' the eigenmodes of $<\mathbf{R}>$, the HPA axis can be trained to galvanized response in the presence of particular forms of perturbation.

This is because, if $<\mathbf{R}>$ has been appropriately determined from regression relations, then the $\lambda_k$ will be a kind of multiple correlation coefficient (e.g. Wallace and Wallace, 2000), so that particular eigenpatterns of perturbation will have greatly amplified impact on the behavior of the HPA axis. If $\lambda = 0$ then perturbation has no more effect than its own magnitude. If, however, $\lambda \to 1$, then the written image of a perturbing psychosocial stressor will have very great effect on the HPA axis. Following Ives (1995), we call a system with $\lambda \approx 0$ *resilient* since its response is no greater than the perturbation itself.

We suggest, then, that learning by the HPA axis is, in fact, the process of tuning response to perturbation. This is why we have written $<\mathbf{R}>$ instead of simply $\mathbf{R}$: The regression matrix is a tunable set of variables.

Suppose we require that $<\lambda>$ itself be a function of the magnitude of excitation, i.e.

$$<\lambda> = f(|<W>|)$$

where $|<W>|$ is the vector length of $<W>$. We can, for example, require the amplification factor $1/(1 - <\lambda>)$ to have a signal transduction form, an inverted-U-shaped curve, for example the signal-to-noise ratio of a stochastic resonance, so that

$$\frac{1}{1 - <\lambda>} = \frac{1/|<W>|^2}{1 + b \exp[1/(2|<W>|)]}.$$

(13)

This places particular constraints on the behavior of the 'learned average' $<\mathbf{R}>$, and gives precisely the typical HPA axis pattern of initial hypersensitization, followed by anergy or 'burnout' with increasing average stress, a behavior that might well be characterized as 'pathological resilience', and may also have evolutionary significance.

Variants of this model permit imposition of cycles of different length, for example hormonal on top of circadian. Typically this is done by requiring a cyclic structure in matrix multiplication, with a new matrix $\mathbf{S}$ defined in terms of a sequential set of the $\mathbf{R}$, having period $m$, so that

$$\mathbf{S}_t \equiv \mathbf{R}_{t+m} \mathbf{R}_{t+m-1} ... \mathbf{R}_t.$$

Essentially one does matrix algebra 'modulo m', in a sense.

In general, while the eigenvalues of such a cyclic system may remain the same, its eigenvectors depend on the choice of phase, i.e. where you start in the cycle. This is a complexity of no small note, and could represent a source of contrast in HPA axis behavior between men and women, beyond that driven by the ten-fold difference in testosterone levels. See Caswell (2001) for mathematical details.

**Obesity as a developmental disorder**

Work by Hirsch (2003) can be interpreted as suggesting that obesity is a developmental disorder with roots in utero or early childhood. Hirsch and others have developed a 'set point' or homeostatic theory of body weight, finding that it is the process which determines that 'set point' which needs examination, rather than the homeostasis itself, which is now fairly well understood. Hirsch concludes that the truly relevant question is not why obese people fail treatment, it is how their level of fat storage became elevated, a matter, he concludes, is probably rooted in infancy and childhood, when strong genetic determinants are shaping a still-plastic organism.

The question we raised earlier regarding the division of sets of possible responses of a cognitive HPA axis into the sets $B_0$ and $B_1$ has special significance in this matter.

Recall that the essential characteristic of cognition in our formalism involves a function $h$ which maps a (convolutional) path $x = a_0, a_1, ..., a_n, ...$ onto a member of one of two disjoint sets, $B_0$ or $B_1$. Thus respectively, either (1) $h(x) \in B_0$, implying no action taken, or (2), $h(x) \in B_1$, and some particular response is chosen from a large repertoire of possible responses. We discussed briefly the problem of defining these two disjoint sets, and suggested that some 'higher order cognitive module' might be needed to identify what constituted $B_0$, the set of 'normal' states. This is because there is no low energy mode for information systems: virtually all states are more or less high energy states involving high rates of information transfer, and there is no way to identify a ground state using the physicist's favorite variational or other extremal arguments.

Suppose that higher order cognitive module, which we might well characterize as a Zero Mode Identification (ZMI), interacts with an embedding language of structured psychosocial stress and, instantiating a Rate Distortion image of that embedding stress, begins to include one or more members of the set $B_1$ into the set $B_0$. If that element of $B_1$ involves a particular mode of HPA axis cortisol/leptin cycle interaction, then the bodymass set point may be reset to a higher value: onset of obesity.

Work by Barker and colleagues suggests that those who develop coronary heart disease (CHD) grow differently from others, both in utero and during childhood. Slow growth during fetal life and infancy is followed by accelerated weight gain in childhood, setting a life history trajectory for CHD, type II diabetes, hypertension, and, of course, obesity. Barker (2002) concludes that slow fetal growth might also heighten the body's stress responses and increase vulnerability to poor living conditions later in life. Thus faulty ZMI function at critical periods in growth could lead to a permanently high body mass set point as a developmental disorder.

Empirical tests of our 'higher order' hypothesis, however, quickly lead into real-world regression models involving the interrelations of measurable biomarkers, behaviors, and so on, requiring formalism much like that used in the section above.

**Recent trajectories of structured stress in the US**

Two powerful and intertwining phenomena of socioeconomic disintegration – deurbanization in the 1970's, and deindustrialization, particularly since 1980 – have combined to profoundly damage many US communities, dispersing historic accumulations of economic, political, and social capital. These losses have had manifold and persisting impacts on both institutions and individuals (e.g. Pappas, 1989; Ullmann, 1988; Wallace and Wallace, 1998). Elsewhere we examined the effect of these policy-driven phenomena on the hierarchical diffusion of AIDS in the US (Wallace et al., 1999). Here we extend that work to their association with obesity, in the context of the causal biological model given above.

By 1980, not a single African-American urban community established before or during World War II remained intact. Many Hispanic urban neighborhoods established after the war suffered similar fates. Virtually all lost considerable housing, population, and economic and social capital either to programs of urban renewal in the 1950's or to policy-related contagious urban decay from the late 1960's through the late 1970's (e.g. Wallace and Wallace, 1998; M. Fullilove, 2004).

Figure 1 exemplifies the process, showing the percent change of occupied housing units in the Bronx section of New York City between 1970 and 1980 by 'Health Area', the aggregation of US Census Tracts by which morbidity and mortality are reported in the city. The South-Central section of the Bronx, by itself one of the largest urban concentrations in the Western world with about 1.4 million inhabitants, had lost between 55 and 80 percent of housing units, most within a five year period. This is a level of damage unprecedented in an industrialized nation short of civil or international war, and indeed can be construed as constituting a kind of covert civil war (Wallace and Wallace, 1998; Duryea, 1978).

Figure 2, a composite index of number and seriousness of building fires from 1959 through 1990 (Wallace et al, 1999; Wallace and Wallace, 1998), illustrates the process of contagious urban decay in New York City which produced that housing loss, affecting large sections of Harlem in Manhattan, and a broad band across the African-American and Hispanic neighborhoods of Northern Brooklyn, from Williamsburg to Bushwick, Brownsville, and East New York. The sudden rise between 1967 and 1968 was stemmed through 1972 by the opening of 20 new fire companies in high fire incidence, minority neighborhoods of the city. Beginning in late 1972, however, some 50 firefighting units were closed and many others destaffed as part of a 'planned shrinkage' program which continued the ethnic cleansing policies of 1950's urban renewal without benefit of either constitutional niceties or new housing construction to shelter the displaced population (e.g. Wallace and Wallace, 1998; Duryea, 1978).

Similar maps and graphs could be drawn for devastated sections of Detroit, Chicago, Los Angeles, Philadelphia, Baltimore, Cleveland, Pittsburgh, Newark, and a plethora of smaller US urban centers, each with its own individual story of active public policy and passive 'benign neglect'.

Figure 1 represents the Bronx part of the spatial distribution of the time integral of figure 2.

Figure 3, using data taken from the US Census, shows the counties of the Northeastern US which lost more than 1000 manufacturing jobs between 1972 and 1987, i.e. the famous 'rust belt'. It is, in its way, an exact parallel to figure 1 in that unionized manufacturing jobs lost remained lost, and their associated social capital and political influence were dispersed. As Pappas (1989) describes, the effects were profound and permanent.

Figure 4, using data from the US Bureau of Labor Statistics, shows the total number of US manufacturing jobs from 1980 to 2001. We define our environmental index of the US national pattern of structured stress to be represented by *the integral of manufacturing job loss after 1980*, i.e. the space between the observed curve and a horizontal line drawn out from the 1980 number of jobs. This is not quite the same as figure 3, which represents a simple net loss between two time periods. We believe that manufacturing job loss at one period continues to have influence at subsequent periods as a consequence of permanently dispersed social and political capital, at least over a 20 year span. Other models, perhaps with different integral weighting functions, are, of course, possible. We use simply

$$D(T) = -\sum_{\tau=1980}^{\tau=T} [M(\tau) - M(1980)]$$

(14)

while a more elaborate treatment might involve something like

$$D(T) = -\int_{\tau_0}^{T} f(T-\tau)M(\tau)d\tau$$

(15)

where $D$ is the deficit, $M(\tau)$ is the number of manufacturing jobs at time $\tau$, and $f(T-\tau)$ is a lagged weighting function.

Figure 5, using data from the Centers for Disease Control (CDC, 2003), shows the percent of US adults characterized as 'obese' according to the Behavioral Risk Factor Surveillance System between 1991 and 2001. This is given as a function of the integrated manufacturing jobs deficit from 1980, again, calculated as a simple negative sum of annual differences from 1980.

The association is quite good indeed, and the theory of the first sections suggests the relation is causal and not simply correlational: Loss of stable working class employment, loss of social and political capital, loss of union influence on working conditions and public policies, deurbanization intertwined with deindustrialization and their political outfalls, all constitute a massive threat expressing itself in population-level patterns of HPA axis-driven metabolism and metabolic syndrome.

Figure 6 extends the analysis to diabetes deaths in the US between 1980 and 1998. It shows the death rate per 100,000 as a function of the cumulative manufacturing jobs deficit from 1980 through 1998. Diabetes deaths are, after a lag, a good index of population obesity. Two systems are evident, before and after 1989, with a 'phase transition' between them probably representing, in Holling's (1973) sense, a change in ecological domain roughly analogous to the sudden eutrophication of a lake progressively subjected to contaminated runoff. This would seem to reflect a the delayed cumulative impacts of both deindustrialization and the deurbanization which became closely coupled with it. Further sudden, marked, upward transitions seem likely if socioeconomic and political reforms are not forthcoming. A parallel paper to this one examines the details of such large-scale 'cognitive' ecosystem resilience, as opposed to the simple Ives model adapted above (Wallace and Wallace, 2003).

It would be useful to compare annual county-level maps of diabetes death rates with those of manufacturing job loss and deurbanization, but such a study would require considerable resources in order to conduct the necessary sophisticated analyses of cross-coupled, lagged, spatial and social diffusion.

### Discussion and conclusions

Current theory clearly identifies 'stress' as critical to the etiology of visceral obesity, the metabolic syndrome, and their pathological sequelae, mediated by the HPA axis and several other physiological subsystems which we have not addressed here.

Both animal and human studies, however, have indicated that not all stressors are equal in their effect: particular forms of domination in animals and lack of control over work activities in humans are well-known to be especially effective in triggering metabolic syndrome and chronic inflammatory coronary lipid deposition.

Our analysis has been in terms of a cognitive HPA axis responding to a highly structured 'language' of psychosocial stress, which we see as literally writing a distorted image of itself onto the behavior of the HPA axis in a manner analogous to learning plateaus in a neural network or to punctuated equilibrium in a simple evolutionary process. The first form of 'phase transition' might be regarded as representing the progression of a normally 'staged' disease. The other could describe certain pathologies characterized by stasis or only slight change, with staging a rare (and perhaps fatal) event.

Psychosocial stress is, for humans, a cultural artifact, one of many such which interact intimately with human physiology. Indeed, much current theory in evolutionary anthropology focuses on the essential (but not unique) role culture plays in human biology (e.g. Durham, 1991; Avital and Jablonka, 2000).

If, as the evolutionary anthropologist Robert Boyd has suggested, "Culture is as much a part of human biology as the enamel on our teeth", what does the rising tide of obesity in the US suggest about American culture and the American system? About 22% of both African-American and Hispanic children are overweight, as compared to about 12% of non-Hispanic whites, and that prevalence is rising across the board (Strauss and Pollack, 2001). This suggests that, while the effects of an accelerating social pathology related to deindustrialization, deurbanization, and loss of democracy may be most severe for ethnic minorities in the US, the larger, embedding, cultural dysfunction has already spread upward along the social hierarchy, and is quickly entraining the majority population as well.

This is an explanation whose policy implications stand in stark contrast to current individual-oriented exhortations about 'taking responsibility for one's behavior' or 'eating less and getting more exercise' (e.g. Hill et al., 2003). The US 'liberal' approach is to mirror the explanations of the failed drug war: People overeat because there's a McDonald's on every street corner, companies market bigger portions, the food they sell is fatty, and so on. In contrast, we find that the fundamental cause of the obesity epidemic over the last twenty years is not television, the automobile, or junk food. All were prominent from the late 1950's into the 1980's without an obesity epidemic. The fundamental cause of the US obesity epidemic is a massive threat to the population caused by continuing deterioration of basic US social, economic, and related structures, in the particular context of a ratcheting of dominance relations resulting from the concentration of effective power within a shrinking elite. This phenomenon is literally writing a life-threatening image of itself onto the bodies of American adults and children. There is already a large and growing literature on other aspects of the sharpening inequalities within the US system (e.g. Wilkinson, 1996 and related material), and our conclusions fit within that body of work.

The basic and highly plieotropic nature of the biological relation between structured psychosocial stress and cognitive physiological systems ensures that 'magic bullet' interventions will be largely circumvented: in the presence of a continuing socioeconomic and political ratchet, 'medical' modalities are likely to provide little more than the equivalent of a choice of dying by hanging or by firing squad.

Effective intervention against obesity in the US is predicated on creation of a broad, multi-level, ecological control program. It is evident that such a program must include redress of the power relations between groups, rebuilding of urban (and, increasingly, suburban) minority communities, and effective reindustrialization. This implies the necessity of a resurgence of the labor union, religious, civil rights, and community-based political activities which have been traditionally directed against cultural patterns of injustice in the past, activities which, ultimately, liberate all.

## Acknowledgments

This work benefited from support under NIEHS Grant I-P50-ES09600-05 and through earlier funding from a Robert Wood Johnson Foundation Investigator Award in Health Policy Research.

## References

Adams F., 2003, the informational turn in philosophy, *Minds and Machines*, 13:471-501.

Allan, J., 1998,Explanatory models of overweight among African American, Euro-American, and Mexican American women,*Western Journal of Nursing Research*, 20:45-66.

Atlan H. and I. Cohen, 1998, Immune information, self-organization and meaning, *International Immunology*, 10:711-717.

Avital E. and E. Jablonka, 2000, *Animal Traditions: behavioral inheritance in evolution*, Cambridge University Press, UK.

Barker D., 2002, Fetal programming of coronary heart disease, *Trends in Endocrinology and Metabolism*, 13:364-372.

Binney J., N. Dowrick, A. Fisher, and M. Newman, 1986, *The theory of critical phenomena*, Clarendon Press, Oxford, UK.

Bjorntorp P., 2001, Do stress reactions cause abdominal obesity and comorbidities?, *Obesity Reviews*, 2:73-86.

Brunner E., M. Marmot, K. Nanchahal, M. Shipley, S. Stansfeld, M. Juneja and K. Alberti, 1997, Social inequality in coronary risk: central obesity and the metabolic syndrome. Evidence from the Whitehall II study, *Diabetologia*, 40:1341-1349.

CDC, 2003, 1991-2001 Prevalence of Obesity Among US Adults, by Characteristics: Behavioral Risk Factor Surveillance System (1991-2001);Self-reported data.

http://www.cdc.gov/nccdphp/dnpa/obesity/trend/prev$_c$har.ht

Chrousos G., 2000, The role of stress and hypothalamic-pituitary-adrenal axis in the pathogenesis of the metabolic syndrome: neuro-endocrine and target tissue-related causes, *International Journal of Obesity and Related Metabolic Disorders*, Suppl. 2:S50-S55.

Cover T. and J. Thomas, 1991, *Elements of Information Theory*, John Wiley Sons, New York.

Caswell H., 2001, *Matrix Population Models*, Aldine.

Dimitrov A. and J. Miller, 2001, Neural coding and decoding: communication channels and quantization, *Computation and Neural Systems*, 12:441-472.

Durham W., 1991, *Coevolution: Genes, Culture, and Human Diversity*, Stanford University Press, Palo Alto, CA.

Duryea P., 1978, Press release dated Friday 27 January, Office of the New York State Assembly Republican Leader, Albany, NY. Copy available from R. Wallace.

Eldredge N., 1985, *Time Frames: The Rethinking of Darwinian Evolution and the Theory of Punctuated Equilibria*, Simon and Schuster, New York.

Feynman R., 1996, *Feynman Lectures on Computation*, Addison-Wesley, Reading, MA.

Fullilove M., 2004, *Root Shock: Upheaval, Resettlement and Recovery in Urban America*, Random House, in press.

Ginsberg-Fellner F., L. Jagendorf, H. Carmel, and T. Harris, 1981, *American Journal of Nutrition*, Overweight and obesity in preschool children in New York City,34:2236-2241.

Gould S., 2002, *The Structure of Evolutionary Theory*, Harvard University Press, Cambridge, MA.

Hill J., H. Wyatt, G. Reed, and J. Peters, 2003, Obesity and the environment: where do we go from here? *Science*, 266:853-858.

Hirsch J., 2002, Obesity: matter over mind?, *Cerebrum*, xx:7-18.

Holling C., 1973, Resilience and stability of ecological systems, *Annual Review of Ecological Systems and Systematics*, 4:1-25.

Ives A., 1995, Measuring resilience in stochastic systems, *Ecological Monographs*, 65:217-233.

Mokdad A., M. Serdula, W. Dietz, B. Bowman, J. Marks, J. Koplan, 1999, The spread of the obesity epidemic in the United States, 1991-1998 *JAMA*, 282:1519-1522.

Pappas G., 1989, *The Magic City*, Cornell University Press, Ithaca, NY.

Rojdestvensky I. and M. Cottam, 2000, Mapping of statistical physics to information theory with applications to biological systems, *Journal of Theoretical Biology*, 202:43-54.

Spiegel K., R. Leproult, and E. Van Cauter, 1999, Impact of sleep debt on metabolic and endocrine function, *The Lancet*, 354:1435-1439.

Strauss R. and H. Pollack, 2001, Epidemic increase in childhood overweight, 1986-1998, *Journal of the American Medical Association*, 286:2845-2848.

Ullmann J., 1988, *The Anatomy of Industrial Decline*, Greenwood-Quorum, Westport, CT.

Wallace D., and R. Wallace, 1998, *A Plague on Your Houses*, Verso Publications, New York

Wallace D. and R. Wallace, 2000, Life and death in Upper Manhattan and the Bronx: toward an evolutionary perspective on catastrophic social change, *Environment and Planning A*, 32:1245-1266.

Wallace R. and R.G. Wallace, 1998, Information theory, scaling laws and the thermodynamics of evolution, *Journal of Theoretical Biology*, 192:545-559.

Wallace R., and R.G. Wallace, 1999, Organisms, organizations, and interactions: an information theory approach to biocultural evolution, *BioSystems*, 51:101-119.

Wallace R., D. Wallace, J. Ullmann, and H. Andrews, 1999, Deindustrialization, inner-city decay, and the diffusion of AIDS in the USA, *Environment and Planning A*, 31:113-139.

Wallace R., 2000, Language and coherent neural amplification in hierarchical systems: renormalization and the dual information source of a generalized stochastic resonance, *International Journal of Bifurcation and Chaos*, 10:493-502.

Wallace R., 2002a, Immune cognition and vaccine strategy: pathogenic challenge and ecological resilience, *Open Systems and Information Dynamics*, 9:51-83.

Wallace R., 2002b, Adaptation, punctuation and rate distortion: non-cognitive 'learning plateaus' in evolutionary process, *Acta Biotheoretica*, 50:101-116.

Wallace R., D. Wallace, and R.G. Wallace, 2003, Toward cultural oncology: the evolutionary information dynamics of cancer, *Open Systems and Information Dynamics*, 10:159-181. Draft available at

http://cogprints.ecs.soton.ac.uk/archive/00002702/

Wallace R. and D. Wallace, 2003, Resilience reconsidered: case histories from disease ecology.

http://www.arxiv.org/abs/q-bio.NC/0310009

Wallace R., 2003, The cognitive homunculus: do tunable languages-of-thought convey adaptive advantage?

http://www.arxiv.org/abs/q-bio.NC/0312003

Wellman N. and B. Friedberg, 2002, Causes and consequences of adult obesity: health, social and economic impacts in the United States, *Asia Pacific Journal of Clinical Nutrition*, Suppl.8:S705-S709.

Wilkinson R., 1996, *Unhealthy Societies: the afflictions of inequality*, Routledge, London and New York.

Wilson, K., 1971, Renormalization group and critical phenomena. I Renormalization group and the Kadanoff scaling picture, *Physical Review B*, 4:3174-3183.

**Figure Captions**

**Figure 1.** Percent change in occupied housing units, Bronx section of New York City, 1970-1980. Notice that large areas lost over half their housing in this period, a degree of destruction unprecedented in an industrialized nation outside of wartime. Similar policy-driven disasters have afflicted most US urban minority communities since the end of World War II.

**Figure 2.** Annual fire damage index New York City, 1959-1990. This is a composite of number and seriousness of structural fires, and is an index of contagious urban decay. Some 20 new fire companies were added to high fire areas between 1969 and 1971, interrupting the process. Fifty firefighting units were closed in or permanently relocated from, high fire areas after November, 1972, allowing contagious urban decay to proceed to completion, producing the conditions of figure 1.

**Figure 3.** The Rust Belt: counties of the Northeastern US which lost 1000 or more manufacturing jobs between 1972 and 1987.

**Figure 4.** Annual number of manufacturing jobs in the US, 1980-2001. Our environmental index of social decay is the integrated loss after the 1980 peak, representing the permanent dispersal of economic, social, and political capital, part of the 'opportunity cost' of a deindustrialization largely driven by the diversion of technical resources from civilian industry into the Cold War (e.g. Ullmann, 1988).

**Figure 5.** 1991-2001 relation between adult obesity in the US and the integrated loss of manufacturing jobs after 1980. We believe manufacturing job loss is an index of permanent decline in social, economic, and political capital which is perceived as, and indeed represents, a serious threat to the well-being of the US population.

**Figure 6.** 1980-1998 relation between US diabetes death rate and integrated loss of manufacturing jobs after 1980. Two systems are evident: before and after 1989. We believe this sudden change represents a nonlinear transition between ecosystem domains which is much like the eutrophication of an increasingly contaminated water body (e.g. Holling, 1973).

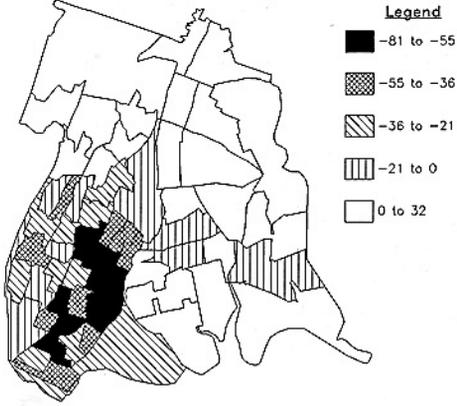

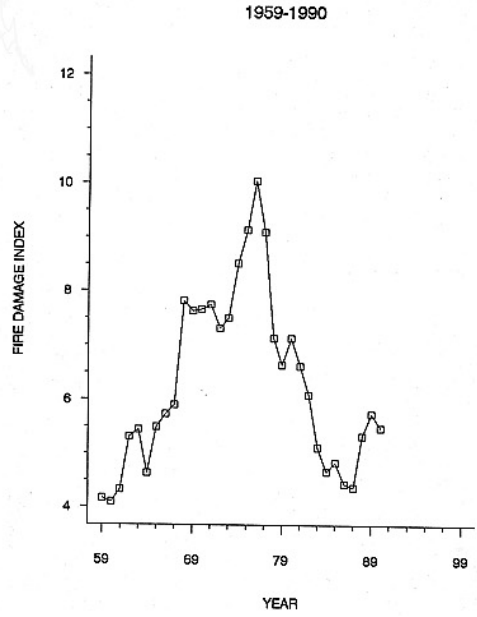

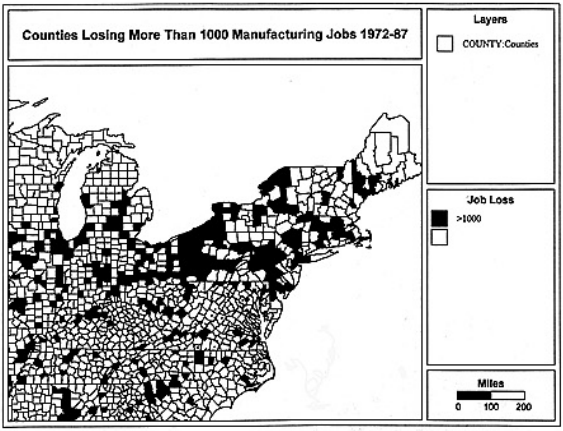

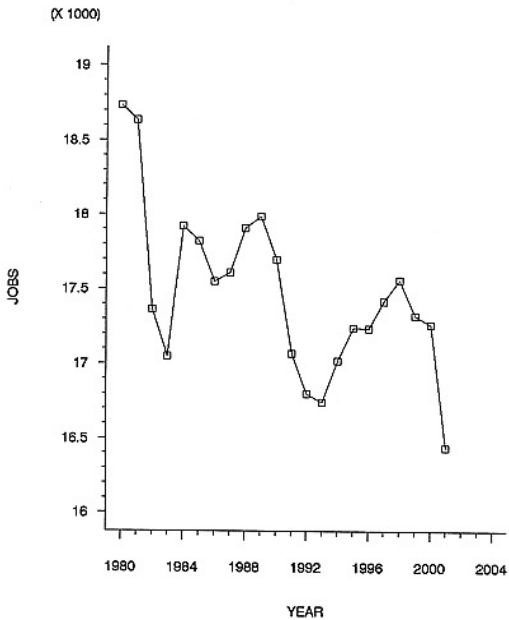

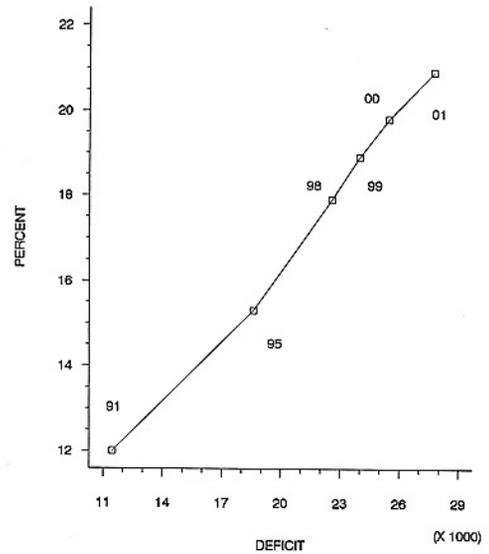

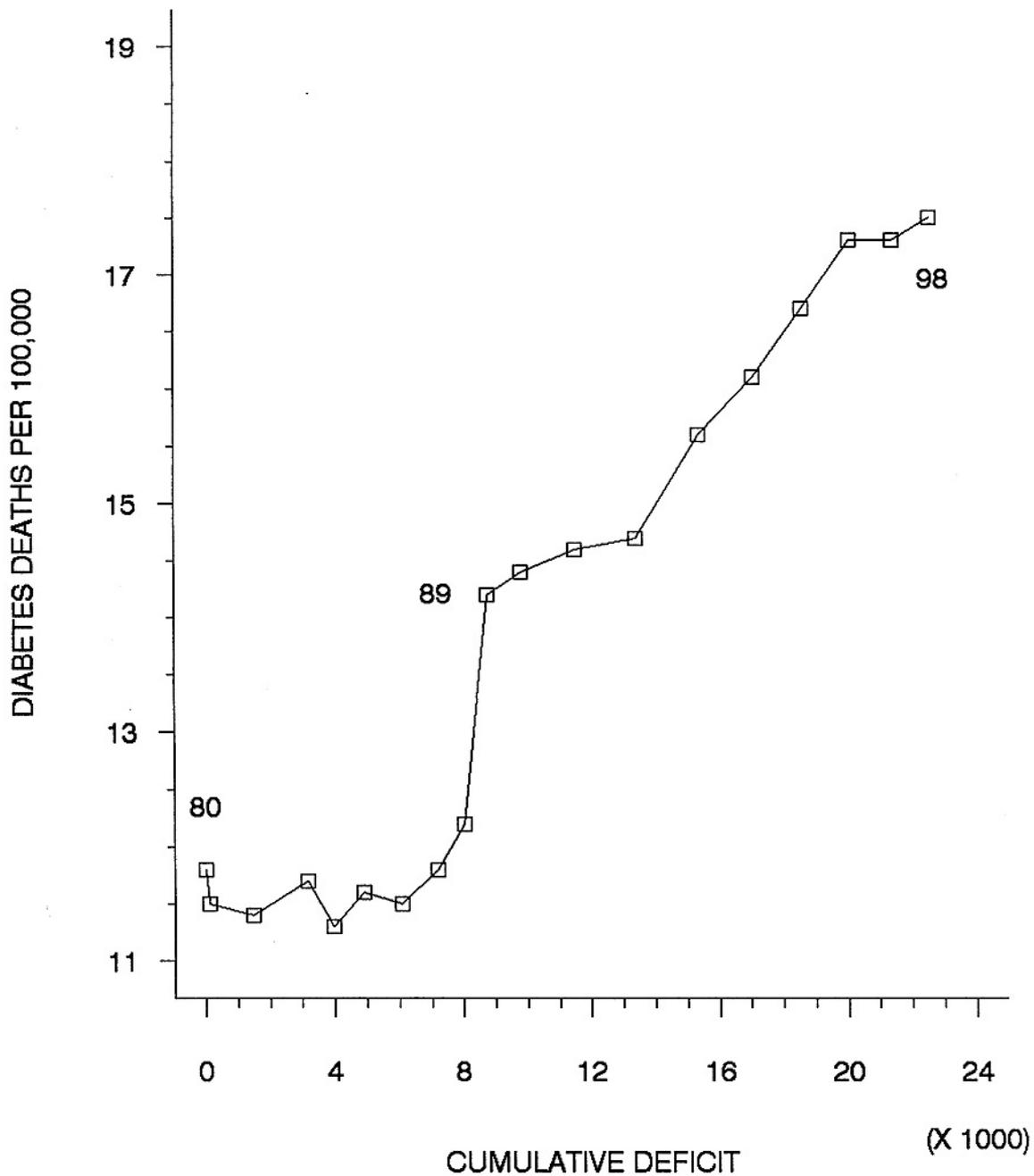